\documentclass[eqsecnum,preprint,prd,aps,nofootinbib]{revtex4}
\usepackage{amsmath,amsfonts}
\usepackage{amssymb}
\usepackage{graphicx}
\usepackage{epsfig}
\newcommand{\be}{\begin{equation}}
\newcommand{\ee}{\end{equation}}
\newcommand{\ba}{\begin{eqnarray}}
\newcommand{\ea}{\end{eqnarray}}
\newcommand{\bc}{\begin{center}}
\newcommand{\ec}{\end{center}}
\def\lvec#1{\vbox{\ialign{##\crcr$\leftarrow$\crcr\noalign{
 \kern-1pt\nointerlineskip}$\hfil\displaystyle{#1}\hfil$\crcr}}}
\def\rvec#1{\vbox{\ialign{##\crcr$\rightarrow$\crcr\noalign{
 \kern-1pt\nointerlineskip}$\hfil\displaystyle{#1}\hfil$\crcr}}}

\newcommand{\defeq}{\stackrel{\rm{def}}{=}}

\makeatletter
\newenvironment{sizeequation}[1]{%
  \skip@=\baselineskip
  #1%
  \baselineskip=\skip@
  \equation
}{\endequation \ignorespacesafterend}
\makeatother

\begin{document}
\begin{center}
\bibliographystyle{article}

{\Large \textsc{On the complete analytic structure of the massive 
gravitino propagator in four-dimensional de Sitter space}}

\end{center}
\vspace{0.4cm}

\author{Giampiero Esposito$^{1}$ \thanks{
Electronic address: giampiero.esposito@na.infn.it}
Raju Roychowdhury$^{2,1}$ \thanks{
Electronic address: raju@na.infn.it}}

\affiliation{
${\ }^{1}$Istituto Nazionale di Fisica Nucleare, Sezione di Napoli,\\
Complesso Universitario di Monte S. Angelo, Via Cintia, Edificio 6, 80126
Napoli, Italy\\
${\ }^{2}$Dipartimento di Scienze Fisiche, Federico II University,\\
Complesso Universitario di Monte S. Angelo, Via Cintia, Edificio 6,
80126 Napoli, Italy}

\vspace{0.4cm}
\date{\today}

\begin{abstract}
With the help of the general theory of the Heun equation, this
paper completes previous work by the authors and other groups on
the explicit representation of the massive gravitino propagator in
four-dimensional de Sitter space. 
As a result of our original contribution,
all weight functions which multiply the geometric
invariants in the gravitino propagator are expressed through Heun
functions, and the resulting plots are displayed and discussed
after resorting to a suitable truncation in the series expansion
of the Heun function. It turns out that there exist two ranges of values 
of the independent variable in which the weight functions can be divided
into dominating and sub-dominating family. 
\end{abstract}

\maketitle
\bigskip
\vspace{2cm}

\section{Introduction}

The investigation of Green functions has always been at the heart of
important developments in quantum field theory and quantum gravity 
\cite{DeWi65}. On the other hand, in recent years, 
developments in cosmology and string theory have
led to renewed interest in supergravity theories in anti-de Sitter
\cite{Witten98} and de Sitter space \cite{Witten01}.

Thus, in our recent paper \cite{raju2}, 
we performed a two-component spinor analysis
of geometric invariants leading to the gravitino 
propagator in four-dimensional
de Sitter spacetime, following the two-spinor language \cite{penrose} 
pioneered by Penrose. In that paper we also wrote down all the 10 different 
weight functions multiplying the invariants which occur in the 
massive gravitino propagator, relying upon the work by Anguelova
and Langfelder \cite{anguelova}.
It was also found there, that algebraically one can write down 
8 weight functions denoted by 
$\alpha, \beta, \gamma, \delta, \varepsilon, \theta, \tau, \omega$ 
in terms of a pair denoted by $(\pi, \kappa)$, in case of de Sitter space. 
Going one step further, we
also expressed $\kappa$ in terms of $\pi$ and $\pi'$
where $\pi$ was defined in this fashion: $\pi (z) = 
\sqrt{z} \, \tilde{\pi} (z)$ and $\tilde{\pi} (z)$
satisfies the Heun differential equation \cite{handbook,heundiff},
whose solutions, denoted by ${\rm Heun}(a,q;b,c,d,e;z)$, 
with properly defined 
arguments, have in general four singular points, i.e. $z_{0}=0,1,a,\infty$. 

In this paper we have explicitly written down, first, the algebraic expression 
of all the 9 weight functions  $\kappa,\alpha, \beta, 
\gamma, \delta, \varepsilon, \theta, \tau, \omega$
in terms of $\pi(z)$ and $\pi'(z)$, where $z$ is defined 
as $z = \cos^2\frac{\mu}{2R}$, with 
$\mu(x,x')$ being the geodesic distance between 
$x$ and $x'$ as defined in \cite{raju2}. Finally, 
we will draw a few two-dimensional plots of these weight 
functions and classify their 
parameter space with respect to $z$ in the region of our choice.

The plan of this paper is as follows. In Sec. II we set up all  
symbols, basically recalling
all relevant definitions of use in this paper from our 
previous one \cite{raju2}.
Sec. III contains the explicit massive spin-3/2 propagator 
in four dimensions with all the ten invariant structures
properly defined, along with the multiplicative weight functions 
written in terms of the $(\pi, \kappa)$ pair. 
In Sec. IV we give a crash course on Heun's 
differential equations and
write down several properties of the Heun function before showing 
that $\tilde{\pi}(z)$
satisfies a Heun equation with properly defined arguments that 
we will list there. Sec. V and the appendix are devoted to build
a dictionary of all the 9 weight functions 
$\kappa,\alpha, \beta, \gamma, \delta, \varepsilon, 
\theta, \tau, \omega$ written in terms of $\pi(z)$ and $\pi'(z)$ 
only, where prime denotes derivative with respect to
$z$ instead of being the derivative with respect to $\mu$,
the geodesic distance function. Then in the last Sec. VI we display 
several two-dimensional plots showing the functional behavior of each 
of the 10 weight functions with respect to $z$. These show that
there exist two ranges of values of $z$ in which the weight functions
can be divided into dominating and sub-dominating family.
Moreover, it appears helpful to
have the result of a lengthy calculation completely worked out. 
Eventually, we give further details on the plots in the section devoted to
concluding remarks. 

In light of recent mathematical developments in \cite{2009}, 
it might be possible to expand the Heun functions
in the gravitino propagator as a combination of finitely or infinitely
many hypergeometric functions, which in turn occur in the more familiar
formulae for bosonic propagators in de Sitter space 
\cite{1987}. Thus, our work
might help relating fermionic and bosonic propagators in 
four-dimensional de Sitter space through special-function techniques,
double-checking the expectations from supersymmetry. 

\section{A review of a few useful definitions}

It has been more than two decades since Allen and co-authors used 
intrinsic geometric objects to calculate correlation functions in maximally
symmetric spaces; their results, here exploited, were presented in two 
papers \cite{allen1,allen2}. In this section we would like to 
review, first, the elementary 
maximally symmetric bi-tensors which have been discussed 
previously by Allen and Jacobson \cite{allen1}. 
More recently, the calculation of 
the spinor parallel propagator 
has been carried out in arbitrary dimension \cite{mueck}.

A maximally symmetric space is a topological manifold of dimension $n$, 
with a metric which has the maximum number of global Killing 
vector fields. This type of space looks exactly the same in every 
direction and at every point. The simplest examples are flat space 
and sphere, each of which 
has $\case{1}{2}n(n+1)$ independent Killing fields.
For $S^n$ these generate all rotations, and for 
$\mathbb{R}^n$ they include both rotations and translations.

We consider a maximally symmetric space of dimension $n$ with constant
scalar curvature $n(n-1)/R^2$. For the space $S^n$, the radius $R$ is
real and positive, whereas for the hyperbolic space $H^n$, $R=il$ with
$l$ positive, and in the flat case, $\mathbb{R}^n$, $R=\infty$. If we  
further consider two points $x$ and $x'$, which can be connected
uniquely by a geodesic, with $\mu(x,x')$ being the geodesic
distance between $x$ and $x'$, then $n^a(x,x')$
and $n^{a'}(x,x')$ are the tangents to the geodesic at $x$ and $x'$, 
and are given in terms of the geodesic distance as follows:
\begin{equation}
\label{ndef} 
n_{a}(x,x') = \nabla_{a}\mu(x,x') \quad \text{and} \quad 
n_{a'}(x,x') = \nabla_{a'} \mu(x,x').
\end{equation}
Furthermore, on denoting
by $g^{a}_{\;b'}(x,x')$ the vector parallel propagator along the
geodesic, one can then write $n^{b'} = -g^{b'}_{\;a} n^a$. Tensors
that depend on two points, $x$ and $x'$, are 
bitensors \cite{Synge}. They may carry unprimed or primed indices that 
live on the tangent space at $x$ or $x'$.

These geometric objects $n^a$, $n^{a'}$ and
$g^a_{\;b'}$ satisfy the following properties \cite{allen1}:
\begin{subequations}
\begin{align}
\label{dn}
 \nabla_a n_b &= A(g_{ab} -n_a n_b), \\
\label{dnprime}
 \nabla_a n_{b'} &= C(g_{ab'} +n_a n_{b'}), \\
\label{dg}
 \nabla_a g_{bc'} &= -(A+C) (g_{ab} n_{c'} +
 g_{ac'} n_b),
\end{align}
\end{subequations}
where $A$ and $C$ are functions of the geodesic distance 
$\mu$ and are given by \cite{allen1} 
\begin{equation}
\label{AC}
 A = \frac1R \cot \frac{\mu}R \quad \text{and} \quad 
 C = -\frac1{R\sin(\mu/R)}, 
\end{equation} 
for de Sitter spacetime and thus they satisfy the relations 
\begin{equation}
\label{ACrel}
dA/d\mu =-C^2, \quad dC/d\mu =-AC \quad \text{and} \quad C^2-A^2
=1/R^2.
\end{equation}
Last, with our convention the covariant gamma matrices satisfy the property
\begin{equation}
\{\Gamma^\mu,\Gamma^\nu\} =2 I g^{\mu\nu}.
\label{(3.5)}
\end{equation}
In our previous work \cite{raju2}, we followed the conventions 
for two-component spinors, 
as well as all signature and curvature conventions, 
of Allen and Lutken \cite{allen2}, and hence we used dotted and 
undotted spinors instead 
of the primed and unprimed ones of Penrose and Rindler  
\cite{penrose}. In our work a primed index indicates instead that
it lives in the tangent space at $x'$, while the unprimed ones live at $x$.
The fundamental object to deal with is the bispinor $D_A^{\;A'}(x,x')$
which parallel transports a two-component spinor 
$\phi^A$ at the point $x$, along the 
geodesic to the point $x'$, yielding a new spinor $\chi^{A'}$ at $x'$, i.e.
\begin{equation}
\label{def}
\chi^{A'}=\phi^{A} \;  D_{A}^{\;A'}(x,x').
\end{equation}
Complex conjugate spinors are similarly transported by the complex conjugate
of $D_A^{\;A'}(x,x')$, which is $\overline{D}_{\dot{A}}^{\;{\dot{A}'}}(x,x')$.
A few elementary properties of $D_A^{\;A'}$ were listed in 
Sec. IV of \cite{raju2}. It is worth mentioning 
that the covariant derivatives of the spinor parallel propagator 
were defined to be
\begin{equation} 
\label{defcovder}
\nabla_{A\dot{A}}D_B^{\;B'}= (A+C)\left[
\frac{1}{2}n_{A\dot{A}}D_B^{\;B'}-n_{B\dot{A}}D_A^{\;B'}\right],
\end{equation}
where $A$ and $C$ are defined in (2.3).

The two basic massive two-point functions for spin-1/2 particle, 
were defined by 
\begin{equation}
\label{defP}
P^{A{\dot{B}}'} \equiv \langle\phi^{A}(x)
\overline{\phi}^{{\dot{B}}'}(x')\rangle = f(\mu)D^A_{\;A'}n^{A'{\dot{B}}'},
\end{equation}
\begin{equation}
\label{defQ}
Q_{\dot{A}}^{{\;\dot{B}}'} \equiv \langle
\overline\chi_{\dot{A}}(x)\overline{\phi}^{{\dot{B}}'}(x')
\rangle = g(\mu)\overline{D}_{\dot{A}}^{\;{\dot{B}'}}.
\end{equation}
and in de Sitter space they turned out to be \cite{raju2}:
\begin{equation}
\label{2ptfn1}
P^{A{\dot{B}}'}_{(F)} = \lim_{\epsilon \to 0^{+}}f_{DS}
(Z+i\epsilon)D^A_{\;A'}n^{A'{\dot{B}}'},
\end{equation}
\begin{equation}
\label{2ptfn2}
Q^{\dot{A}{\dot{B}}'}_{(F)} = \lim_{\epsilon \to 0^{+}}
g_{DS}(Z+i\epsilon)\overline{D}^{\dot{A}{\dot{B}}'},
\end{equation}
where $(F)$ stands for the Feynman Green functions with 
$f_{DS}$ and $g_{DS}$ defined in this fashion:
\begin{equation}
\label{soln1}
f_{DS} = N_{DS}(1-Z)^{1/2}F(a,b;c;Z),
\end{equation}
\begin{equation}
\label{soln2}
g_{DS}= -iN_{DS}2^{-3/2}m|R|Z^{1/2}F(a,b;c+1;Z).
\end{equation}
Moreover, after doing some algebra one can rewrite the final answer 
for the constant $N_{DS}$ as
\begin{equation}
\label{NDSfinalform}
N_{DS} = \frac{-i|Rm|(1-m^{2}R^{2})}{8\sqrt{2}\pi|R|^{3}\sinh\pi|Rm|}.
\end{equation}
We also note that $F(a,b;c;Z)$ and $F(a,b;c+1;Z)$ 
are two independent solutions of the Hypergeometric 
equation \cite {abra,erdelyi} :
\begin{equation}
\label{hypergeometric}
H(a,b,c;Z)w(Z) = 0,
\end{equation}
where $H(a,b,c)$ is the hypergeometric operator
\begin{equation}
\label{hypergeoop}
H(a,b,c;Z) = Z(1-Z)\frac{d^2}{dZ^2}+[c-(a+b+1)Z]\frac{d}{dZ}-ab.
\end{equation}

\section{Massive spin-3/2 propagator}

In this section we consider the propagator of the massive 
spin-3/2 field. Let us denote the gravitino field by 
$\Psi^{\alpha}_{\lambda} (x)$. In a maximally symmetric state 
$|\, s \rangle$ the propagator is 
\begin{equation}
\label{correlator}
S^{\alpha \beta^{\prime}}_{\lambda \nu^{\prime}} 
(x, x^{\prime}) = \langle s\,| \Psi^{\alpha}_{\lambda} (x) 
\Psi^{\beta^{\prime}}_{\nu^{\prime}} (x^{\prime}) |\,s \rangle .
\end{equation}
The field equations imply that $S$ satisfies 
\begin{equation}
\label{EoM} 
(\Gamma^{\mu \rho \lambda} D_{\rho} - m \, 
\Gamma^{\mu \lambda})^{\alpha}{}_{\gamma} 
S_{\lambda \nu^{\prime}}{}^{\gamma}{}_{\beta^{\prime}} =
\frac{\delta (x-x^{\prime})}{\sqrt{-g}} 
g^{\mu}{}_{\nu^{\prime}} \, \delta^{\alpha}{}_{\beta^{\prime}}. 
\end{equation}

\subsection{The ten gravitino invariants}

It is very convenient to decompose the gravitino propagator in terms 
of independent structures
constructed out of $n_\mu, n_{\nu'}, g_{\mu\nu'}$ and 
$\Lambda^\alpha_{~\beta'}$ \cite{raju2}.
Thus, the propagator can be written in geometric way following Anguelova 
et al. \cite{anguelova} (see also \cite{Basu}):
\begin{eqnarray}
\label{ansatz}
S_{\lambda \nu^{\prime}}{}^{\alpha}{}_{\beta^{\prime}} 
&=& \alpha (\mu) \, g_{\lambda \nu^{\prime}} 
\Lambda^{\alpha}{}_{\beta^{\prime}} + 
\beta (\mu) \, n_{\lambda} n_{\nu^{\prime}} 
\Lambda^{\alpha}{}_{\beta^{\prime}} + 
\gamma (\mu) \, g_{\lambda \nu^{\prime}} (n_{\sigma} 
\Gamma^{\sigma} \Lambda)^{\alpha}{}_{\beta^{\prime}} \nonumber \\ 
&& + \delta (\mu) \, n_{\lambda} n_{\nu^{\prime}} (n_{\sigma} 
\Gamma^{\sigma} \Lambda)^{\alpha}{}_{\beta^{\prime}} + 
\varepsilon (\mu) \, n_{\lambda} (\Gamma_{\nu^{\prime}} 
\Lambda)^{\alpha}{}_{\beta^{\prime}} + 
\theta (\mu) \, n_{\nu^{\prime}} (\Gamma_{\lambda} 
\Lambda)^{\alpha}{}_{\beta^{\prime}} 
\nonumber \\ 
&& + \tau (\mu) \, n_{\lambda} (n_{\sigma} \Gamma^{\sigma} 
\Gamma_{\nu^{\prime}} \Lambda)^{\alpha}{}_{\beta^{\prime}} 
+ \omega(\mu)\, n_{\nu^{\prime}} (n_{\sigma} \Gamma^{\sigma} 
\Gamma_{\lambda} \Lambda)^{\alpha}{}_{\beta^{\prime}} 
\nonumber \\ 
&& + \pi (\mu) \, (\Gamma_{\lambda} 
\Gamma_{\nu^{\prime}} \Lambda)^{\alpha}{}_{\beta^{\prime}}  
+ \kappa (\mu)\, (n_{\sigma} \Gamma^{\sigma} 
\Gamma_{\lambda} \Gamma_{\nu^{\prime}} \Lambda)^{\alpha}{}_{\beta^{\prime}}. 
\end{eqnarray}

\subsection{The weight functions multiplying the invariants}

A rather tedious but straightforward calculation gives a system of 
$10$ equations for the $10$ coefficient functions $\alpha, ..., \kappa$ 
in (\ref{ansatz}) as found in (see equations (3.6)-(3.15) 
in \cite{anguelova}). It was also found there that one 
can easily express the algebraic solutions for 
$\alpha, \beta, \gamma, \delta, \varepsilon, \theta, \tau, \omega$ 
in terms of the $(\pi, \kappa)$ pair in case of de Sitter space, i.e.
(hereafter we set $n=4$ in the general formulae of \cite{anguelova},
since only in the four-dimensional case the two-component-spinor
formalism can be applied)
\begin{eqnarray}
\label{alg}
\omega &=& \frac{2mC \kappa + ((A+C)^{2}-m^2) \pi}
{(m^{2}+R^{-2})}, \nonumber\\
\theta &=& \frac{((A-C)^{2}-m^{2}) \kappa - 2mC \pi}
{(m^{2}+R^{-2})}, \nonumber\\
\tau &=& \frac{2mC \kappa + ((A+C)^{2}-m^2) \pi}
{(m^{2}+R^{-2})}, \nonumber\\
\varepsilon &=& \frac{-([(A-C)^2 + 2/R^2] +m^2) 
\kappa + 2mC \pi}{(m^{2}+R^{-2})}, \nonumber\\
\alpha &=& - \tau - 4\pi , \nonumber\\
\beta &=& 2 \omega , \nonumber\\
\gamma &=& \varepsilon - 2 \kappa , \nonumber\\ 
\delta &=& 2\varepsilon + 4 (\kappa -\theta) , 
\end{eqnarray}
where we have used the relation $C^2 - A^2 = 1/R^2$. 

Furthermore, from (\ref{alg}) we can immediately see that
\begin{equation}
\label{Sym}
\tau = \omega \qquad {\rm and} \qquad \varepsilon + \theta 
= - 2 \kappa . 
\end{equation}

\section{Heun's differential equation: a primer}

The canonical form of the general Heun differential equation 
is given by (\cite{heundiff}, \cite{kamke})
\begin{equation}
\label{ae}
{{{d^2}y}\over{dz^2}}+\left({{\gamma}\over{z}}+{{\delta}\over{z-1}}
+{{\epsilon}\over{z-a}}\right)
{{dy}\over{dz}}+{{{\alpha}{\beta}z-q}
\over{z(z-1)(z-a)}}y=0
\end{equation}
The four regular singular points of the equation are located 
at $z=0,1,a,\infty$.
Here $a\in\mathbb{C}$, the location of the fourth singular point,
is a parameter~($a\neq0,1$), and
$\alpha,\beta,\gamma,\delta,\epsilon\in\mathbb{C}$ are exponent-related
parameters. 

The solution space of the Heun differential equation is specified 
uniquely by the following Riemann $P$-symbol:
\begin{sizeequation}
{\small}
\label{eq:Psymbol}
P\left\{
\begin{array}{ccccc}
0&1&d&\infty& \\
0&0&0&\alpha& ;z \\
1-\gamma&1-\delta&1-\epsilon&\beta& 
\end{array}
\right\}.
\end{sizeequation} 
This does not uniquely specify the equation and its solutions, since it
omits the accessory parameter~$q\in\mathbb{C}$. The exponents are
constrained by
\begin{equation}
\label{eq:Pconstraint}
\alpha+\beta-\gamma-\delta-\epsilon+1 = 0.
\end{equation}
This is a special case of Fuchs's relation, according to which the sum of
the $2n$~characteristic exponents of any second-order Fuchsian equation
on~$\mathbb{CP}^1$ with $n$~singular points must equal
$n-2$~\cite{Poole36}.

There are $2\times4=8$ local solutions of (4.1) in all: two per
singular point. If $\gamma$~is not a nonpositive integer, the solution
at~$z=0$ belonging to the exponent zero will be analytic. When normalized
to unity at~$z=0$, it~is called the local Heun function, and is denoted
$Hl(a,q;\alpha,\beta,\gamma,\delta;z)$~\cite{heundiff}. It is the sum
of a Heun series, which converges in a neighborhood
of~$z=0$~\cite{heundiff,Snow52}. In general,
$Hl(a,q;\alpha,\beta,\gamma,\delta;t)$ is not defined when $\gamma$~is a
nonpositive integer.

If $\epsilon=0$ and $q=\alpha\beta d$, the Heun equation loses a singular
point and becomes a hypergeometric equation. Similar losses occur if
$\delta=0$, $q=\alpha\beta$, or $\gamma=0$,~$q=0$. This paper will exclude
the case when the Heun equation has fewer than four 
singular points. The case, in which the
solution of (4.1) can be reduced to 
quadratures, will also be ruled out.
If $\alpha\beta=0$ and~$q=0$, the Heun equation (4.1) is said to
be trivial. Triviality implies that one of the exponents at~$z=\infty$ is
zero (i.e., $\alpha\beta=0$), and is implied by absence of the singular
point at~$z=\infty$ (i.e., $\alpha\beta=0$, $\alpha+\beta=1$, $q=0$).

\subsection{Reducing Heun to hypergeometric}

The transformation to Heun ($\mathfrak{H}$) or hypergeometric 
($\mathfrak{h}$) of a linear
second-order Fuchsian differential equation with singular points at
$z=0,1,d,\infty$ (resp.\ $z=0,1,\infty$), and with arbitrary exponents, is
accomplished by certain linear changes of the dependent variable, called
F-homotopies (see~\cite{erdelyi} and~\cite[\S\,{A}2 and
Addendum,~\S\,1.8]{heundiff}.)  If~an equation with singular points
at~$z=0,1,a,\infty$ has dependent variable~$u$, carrying~out the
substitution $\tilde u(z)=z^{-\rho}(z-1)^{-\sigma}(z-a)^{-\tau} u(t)$ will
convert the equation to a new one, with the exponents at~$z=0,1,d$ reduced
by~$\rho,\sigma,\tau$ respectively, and those at~$z=\infty$ increased by
$\rho+\sigma+\tau$.  By~this technique, one exponent at each finite
singular point can be shifted to zero.

In fact, the Heun equation has a group of F-homotopic automorphisms
isomorphic to~$({\mathbb Z}_2)^3$, since at each of $z=0,1,a$, the
exponents~$0,\zeta$ can be shifted to~$-\zeta,0$, i.e., to~$0,-\zeta$.
Similarly, the hypergeometric equation has a group of F-homotopic
automorphisms isomorphic to $({\mathbb Z}_2)^2$. These groups act on the
$6$~and~$3$-dimensional parameter spaces, respectively. For example, one
of the latter actions is $(a,b;c)\mapsto(c-a,c-b;c)$, which is induced by
an F-homotopy at~$z=1$. From this F-homotopy follows Euler's
transformation~\cite[\S\,2.2]{Andrews99}
\begin{equation}
\label{eq:flip}
{}_2F_1(a,\,b;\,c;\,z)= (1-z)^{c-a-b}{}_2F_1(c-a,\,c-b;\,c;\,z),
\end{equation}
which holds because ${}_2F_1$~is a local solution at~$z=0$, 
rather than at $z=1$.
If the singular points of the differential equation are arbitrarily placed,
transforming it to the Heun or hypergeometric equation will require a
M\"obius (i.e., projective linear or homographic) transformation, which
repositions the singular points to the standard locations.  A~unique
M\"obius transformation maps any three distinct points in~$\mathbb{CP}^1$
to any other three; but the same is not true of four points, which is why
($\mathfrak{H}$)~has the singular point~$a$ as a free parameter.

\subsection{The cross-ratio orbit}
\label{subsec:crossratio}
The characterization of Heun equations that can be reduced to the
hypergeometric equation will employ the cross-ratio orbit of
$\{0,1,d,\infty\}$, defined as follows. If $A,B,C,D\in\mathbb{CP}^1$ are
distinct, their cross-ratio is
\begin{equation}
(A,B;C,D)\defeq
\frac{(C-A)(D-B)}{(D-A)(C-B)}\in\mathbb{CP}^1\setminus\{0,1,\infty\},
\end{equation}
which is invariant under M\"obius transformations.  Permuting $A,B,C,D$
yields an action of the symmetric group~$S_4$
on~$\mathbb{CP}^1\setminus\{0,1,\infty\}$. The cross-ratio is invariant
under interchange of $A,B$ and~$C,D$, and also under simultaneous
interchange of the two points in each pair. 
Thus, each orbit contains no more
than $4!/4=6$ cross-ratios. The possible actions of~$S_4$
on~$s\in\mathbb{CP}^1\setminus\{0,1,\infty\}$ are generated by
$s\mapsto1-s$ and $s\mapsto 1/s$, and the orbit of~$s$ comprises
\begin{equation}
s,\quad 1-s,\quad 1/s,\quad 1/(1-s),\quad s/(s-1),\quad (s-1)/s,
\end{equation}
which may not be distinct.  This is called the cross-ratio orbit of~$s$;
or, if~$s=(A,B;\allowbreak C,D)$, the cross-ratio orbit of the unordered
set $\{A,B,C,D\}\subset\mathbb{CP}^1$.  Two sets of distinct points
$\{A_i,B_i,C_i,D_i\}$ ($i=1,2$) have the same cross-ratio orbit iff they
are related by a M\"obius transformation.

\subsection{Reminder of some of the properties of Heun's function}

Our aim will be to find an integral representation of the Heun 
function as a Frobenius' solution of the Heun equation, given 
in another form as follows \cite{heundiff}:
\begin{eqnarray} 
\label{Heunnew}
&&z(z-1)(z-a)y^{\prime \prime}(z) + \left\{\gamma (z-1)(z-a)
+\delta z(z-a)+ \epsilon z(z-1)\right\} y^{\prime} (z) \nonumber \\
&&+ (\alpha\beta\, z-q) y(z) = 0,
\end{eqnarray}
The Frobenius' solution, noted  $Hl(a,q;\alpha,\beta,\gamma,\delta;z)$ 
is the entire solution defined for the exponent zero at the 
point $z=0$. It admits the power series expansion
\begin{equation}
\label{Heunseries}
Hl(a,q;\alpha,\beta,\gamma,\delta;z) \equiv 
\sum_{n=0}^{\infty} c_{n}z^{n},
\end{equation} 
with $|z|<1$ and $c_{0}=1$, $c_{1}=\frac{q}{\gamma a}$ and 
$\gamma \neq0,-1,-2,.....$

The recursion relation is as follows:
\begin{eqnarray} 
\label{Recursion}
&&a(n+2)(n+1+\gamma)c_{n+2} \nonumber\\
&&= \Bigr[q+(n+1)(\alpha + \beta - \delta +(\gamma 
+ \delta -1)a)+(n+1)^{2}(a+1)\Bigr]c_{n+1}\nonumber \\
&& - (n+\alpha)(n+\beta)c_{n}= 0   \;\;\;\;\; n\geq 0.
\end{eqnarray}
The function $Hl(a,q;\alpha,\beta,\gamma,\delta;z)$ is normalised 
with the relation
\begin{equation}
Hl(a,q;\alpha,\beta,\gamma,\delta;0)=1.
\end{equation}
It admits the following important particular cases 
(\cite{heundiff}, p9, formula(1.3.9)):
\begin{eqnarray}
\label{properties}
Hl(1,\alpha\beta;\alpha,\beta,\gamma,\delta;z) 
= {}_2F_1(\alpha,\beta,\gamma;\,z) \;\;\;\;\forall 
\delta \in\mathbb{C}\nonumber\\
Hl(0,0;\alpha,\beta,\gamma,\delta;z) = {}_2F_1(\alpha,\beta,\alpha
+\beta-\delta+1;\,z) \;\;\;\; \forall \gamma \in\mathbb{C}\nonumber\\
Hl(a,a\alpha\beta;\alpha,\beta,\gamma,\alpha
+\beta-\gamma+1;z) = {}_2F_1(\alpha,\beta,\gamma;\,z),\nonumber\\
\end{eqnarray}
where ${}_2F_1(\alpha,\beta,\gamma;\,z)$ is the usual notation 
for the Gauss hypergeometric function.

\subsection{Application of Heun's equation to our problem} 

Finally we come to the punch line, why do we need these all and 
how does the Heun equation indeed find an application to our problem?
The answer to this goes along the following line:
On using (\ref{Sym}) the differential equations for $\kappa$ and $\pi$, the 
equations (3.14) and (3.15) of \cite{anguelova}, acquire the form
\begin{eqnarray} 
\label{kp}
-(A+C) \theta + \kappa^{\prime} + \frac{1}{2} (A-C)  
\kappa + m \pi &=& 0 , \nonumber \\
(C-A) \omega + \pi^{\prime} + \frac{1}{2} (A+C) \pi 
+ m \kappa &=& 0 ,
\end{eqnarray}
where $\theta$ and $\omega$ are given in (\ref{alg}). Clearly one can 
solve algebraically the second equation for $\kappa$. By differentiating 
the result one obtains also $\kappa^{\prime}$ in terms of 
$\pi$, $\pi^{\prime}$ and $\pi^{\prime \prime}$, 
and substitution of these in the 
first equation yields a second order ODE for $\pi (\mu)$.
Now let us look at the system (\ref{kp}) in case of de Sitter 
spacetime. On inserting $A$ and $C$ from (\ref{AC}) below 
and passing to the globally defined variable $z = \cos^2
\frac{\mu}{2R}$ (see Sec. III), we obtain the following 
differential equation for $\pi$:
\begin{equation}
\label{pisol}
\left[P_{2}\frac{d^2}{dz^2}+ P_{1}\frac{d}{dz}+ P_{0}\right]\pi = 0,
\end{equation}
where $P_{2}$ in (\ref{pisol}) is a quartic polynomial in $z$, i.e.
\begin{equation}
\label{P0} 
P_{2} = 4 \left[m^{2} R^{2}+1 \right] z^4 
-4(2 m^{2} R^{2}+3) z^3
+4(m^{2}R^{2}+2)z^{2}.
\end{equation}
Similarly, $P_{1}$ in (\ref{pisol}) is a cubic polynomial in $z$,
\begin{equation} 
\label{P1}
P_{1} = 16 \left[m^{2}R^{2}+1\right] z^3 
-12 \left[2m^{2}R^{2}+5 \right]z^2 
+ 8 \left(m^{2}R^{2}+2\right) z.
\end{equation}
Last, $P_{0}$ in (\ref{pisol}) is a quadratic polynomial in $z$, i.e.
\begin{equation} 
\label{P2}
P_{0} = \left(4m^{4}-19m^{2}
+32 m^{2}R^{2}+9\right) z^{2} 
- \left(4m^{4}-14m^{2}+32m^{2}R^{2}+21\right)z
-3m^{2}R^{2}-6.
\end{equation}
On making the substitution $\pi (z) = \sqrt{z} \, \tilde{\pi} (z)$, 
(\ref{pisol}) becomes an equation of the type
\begin{eqnarray} 
\label{Heun}
&&z(z-1)(z-a)y^{\prime \prime}(z) + \left\{ (b+c+1)z^2 -
\left[b+c+1+a(d+e)-e\right]z +ad
\right\} y^{\prime} (z) \nonumber \\
&&+ (bc\, z-q) y(z) = 0.
\end{eqnarray}
Written in canonical form it reads as follows:
\begin{equation}
\label{aeapplication}
{{{d^2}y}\over{dz^2}}+\left({d\over{z}}+{e\over{z-1}}
+{(b+c+1)-(d+e)\over{z-a}}\right){{dy}\over{dz}}
+{{bc z-q}\over{z(z-1)(z-a)}}y=0,
\end{equation}
where the parameters in (\ref{aeapplication}) take the values
\begin{eqnarray} 
a &=& \frac{(m^{2}R^{2}+2)}{(m^{2}R^{2}+1)}, \nonumber\\
b &=& 2+imR, \nonumber\\
c &=& 2-imR, \nonumber\\
d &=& e = 3, \nonumber\\
q &=& -\frac{(m^{4}R^{4}+7m^{2}R^{2}+10)}{(m^{2}R^{2}+1)}.
\label{(6.40)}
\end{eqnarray}
The equation (\ref{Heun}) is known as Heun's differential equation 
\cite{handbook,heundiff}.
Its solutions, here denoted by ${\rm Hl}(a,q;b,c,d,e;z)$, have
in general four singular points as we said before, i.e.
$z_{0}=0,1,a,\infty$. Near each singularity the function behaves as 
a combination of two terms that are powers of $(z-z_0)$ with the 
following exponents:
$\{0, 1-d\}$ for $z_0 = 0$, $\{0, 1-e\}$ for $z_0=1$, $\{0, d+e-b-c\}$ 
for $z_0=a$, and $\{b,c\}$ (that is, 
$z^{-b}$ or $z^{-c}$) for $z\to \infty$.

We now insert into the second of Eq. (4.12) the first of Eq. (3.4),
finding eventually
\begin{equation}
\label{kappaform}
\kappa=f^{-1} \left \{ \left[(A-C)((A+C)^{2}-m^{2})
-{1\over 2}(A+C)(m^{2}+R^{-2})\right] \pi
-(m^{2}+R^{-2})\pi' \right \},
\end{equation}
where
\begin{equation}
f \equiv m (m^{2}+R^{-2}+2C(C-A)),
\end{equation}
and $\pi$ and $\pi'$ are meant to be expressed through the Heun function
${\rm Hl}(a,q;b,c,d,e;z)$. Eventually, we will show in the next 
section that all weight functions can be
therefore expressed through such Heun function. The material covered
in the present section and in the previous two is not new, 
and most of it is appropriate only for a 
physics-oriented choice of four-dimensional de Sitter space.

\section{Dictionary of weight functions for the gravitino propagator}

Here we will explicitly list all the weight functions as  
functions of $z = \cos^2\frac{\mu}{2R}$, in order to analyze their 
qualitative behavior as a function of $z$ and de Sitter radius 
$R$ in the next section. Let us recall a few definitions in de Sitter 
space, where $A$ and $C$ are functions of the geodesic distance 
$\mu$ and are given by \cite{allen1} 
\begin{equation}
\label{AC}
A = \frac1R \cot \frac{\mu}R \quad \text{and} \quad 
C = -\frac1{R\sin(\mu/R)}, 
\end{equation}
Since all other weight functions $\alpha, \beta, \gamma, \delta, 
\varepsilon, \theta, \tau, \omega$ 
can be written in terms of the $(\pi, \kappa)$ pair, and in the last 
section we have seen $\kappa$ can also be expressed in a form 
like (\ref{kappaform}), it is evident that all other 9 weight 
functions including $\kappa$, i.e. $\alpha, \beta, \gamma, \delta, 
\varepsilon, \theta, \tau, \omega,\kappa$ can be expressed in terms 
of $\pi(\mu)$ and $\pi'(\mu)$ only.

We can also express $\pi$ as a function of $z$ and $R$ only as 
$\pi = \pi(z)=\pi(\mu= \pm 2R {\rm cos}^{-1} \sqrt{z})$. 
Similarly, by using a few of the familiar trigonometric identities, one 
can transform $\pi'(\mu)$ as 
\begin{equation}
\label{piprimez}
\pi'(\mu) = \mp\frac{1}{R}\sqrt {z(1-z)} \pi'(z).
\end{equation}
One can also write down the expressions of $(A+C)$ and $(A-C)$ in 
terms of $z$ and $R$ only as follows:
\begin{eqnarray}
\label{ApmC}
&&A+C = -\frac{1}{R} \sqrt{\frac{1-z}{z}}, \nonumber\\
&&A-C = \frac{1}{R} \sqrt{\frac{z}{1-z}}.
\end{eqnarray}
Another function appearing quite frequently in our evaluation of all 
the weight functions is $f$, which can be also expressed as a function 
of $z$ and $R$ only as follows:
\begin{equation}
\label{f}
f = m(m^{2}+R^{-2}+R^{-2}(1-z)^{-1}).
\end{equation} 
Now we start by listing all the weight functions in terms of $\pi(z)$ 
and $\pi'(z)$, bearing in mind that
\begin{equation}
\label{pifinalform}
\tilde{\pi} (z) = {\rm Hl}(a,q;b,c,d,e;z),
\end{equation}
\begin{equation}
\label{pitildefinalform}
\pi (z) = \sqrt{z}\;\;{\rm Hl}(a,q;b,c,d,e;z),
\end{equation}
where ${\rm Hl}(a,q;b,c,d,e;z)$ is the Heun function with arguments 
as defined before. One has therefore the lengthy formulae
for all other weight functions
written down in Eqs. (A1)--(A8) of the appendix.

\section{Qualitative behaviors of the weight functions}

Now using the series expansion (\ref{Heunseries}) 
defined before one can numerically 
study the behavior of each weight function, by taking the first 
10 terms of the infinite series (4.8). Indeed, dealing with an
infinite number of terms is impossible, and one has therefore to
resort to approximations, by truncating such a series. On taking 
less than 10 terms, we have found minor departures from the pattern
outlined below in figures 1 to 9, whereas on taking 15 terms, the
pattern in such figures is essentially confirmed.     

We draw for example all these 
weight functions in a two-dimensional plot vs $z$,
in the range (0,1). The plots, which also include  
${\tilde \pi}(z)$, are as follows.
\begin{figure}[!h]
\centerline{\hbox{\psfig{figure=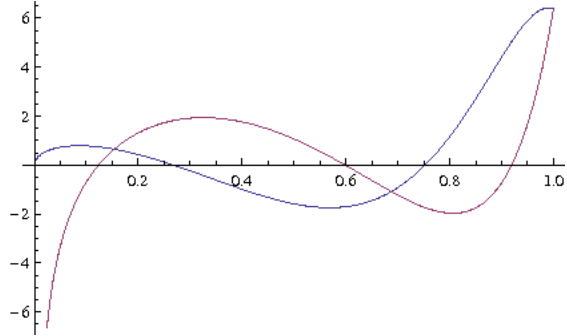,width=0.45\textwidth}}}
\caption{Two-dimensional plot of the weight function $\alpha(z)$.
The curve has two branches, depending on whether one takes 
the $+$ or $-$ sign in (A1). One branch of $\alpha$ cuts the horizontal
$z$-axis at the points $z=0.25, 0.75$, whereas the other branch
of $\alpha$ cuts the horizontal axis at the points
$z=0.12,0.6,0.92$. Both branches approach the vertical axis,
the first one cuts it near the value $0.5$, while the other has
a vertical asymptote at $z=0.02$. The two branches intersect each other
at $z=0.15,0.69,1$; at these points the function $\alpha(z)$ becomes
single-valued.}
\end{figure}
\begin{figure}[!h]
\centerline{\hbox{\psfig{figure=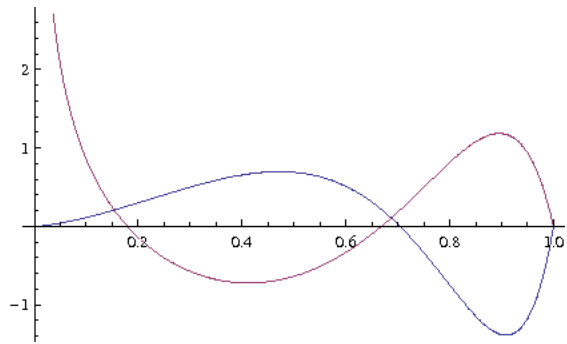,width=0.45\textwidth}}}
\caption{Two-dimensional plot of the weight function $\beta(z)$.
The curve has two branches, depending on whether one takes 
the $+$ or $-$ sign in (A2). One branch of $\beta$ cuts the horizontal
$z$-axis at the points $z=0,0.7,1$, whereas the other branch
of $\beta$ cuts the horizontal axis at the points
$z=0.18,0.67,1$. The first branch never cuts the vertical axis, 
while the other has 
a vertical asymptote at $z=0.05$. The two branches intersect each other
at $z=0.15,0.69,1$; at these points the function $\beta(z)$ becomes
single-valued.}
\end{figure}
\begin{figure}[!h]
\centerline{\hbox{\psfig{figure=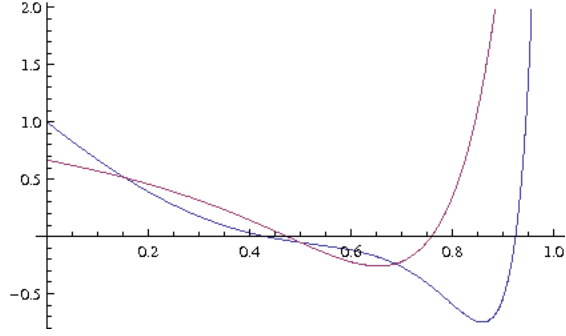,width=0.45\textwidth}}}
\caption{Two-dimensional plot of the weight function $\gamma(z)$.
The curve has two branches, depending on whether one takes 
the $+$ or $-$ sign in (A3). The first branch of 
$\gamma$ cuts the horizontal $z$-axis at the points $z=0.43,0.92$  
and the vertical axis at $1$, and then it has a vertical asymptote 
at $z=0.95$. The second branch
of $\gamma$ cuts the horizontal axis at the points
$z=0.47,0.76$ and the vertical axis at $0.67$, and then it
has a vertical asymptote at $z=0.87$.
The two branches intersect each other
at $z=0.15,0.5,0.69$; at these points the function $\gamma(z)$ becomes
single-valued. The first branch of $\gamma$ has an absolute
minimum, of negative sign, at $z=0.85$.}
\end{figure}
\begin{figure}[!h]
\centerline{\hbox{\psfig{figure=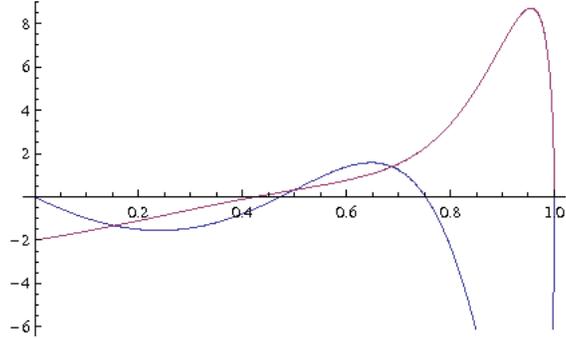,width=0.45\textwidth}}}
\caption{Two-dimensional plot of the weight function $\delta(z)$.
The first branch cuts the horizontal axis at $z=0,0.47,0.75$, 
and the second branch cuts the horizontal axis at $z=0.45,1$.
While the first branch never cuts the vertical axis, the second one 
cuts it at $-2$. The two branches intersect each other at 
$z=0.15,0.5,0.69$, where $\delta$ becomes single-valued.
The first branch has a vertical asymptote at $z=0.85$, whereas
the second one does have the same at $z=1$.}
\end{figure}
\begin{figure}[!h]
\centerline{\hbox{\psfig{figure=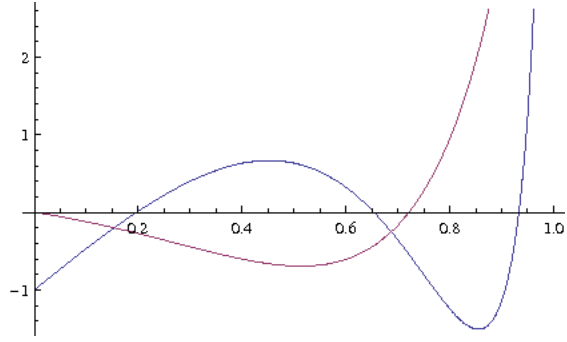,width=0.45\textwidth}}}
\caption{Two-dimensional plot of the weight function $\varepsilon(z)$.
The first branch cuts the horizontal axis at $z=0.2,0.65,0.93$
and cuts the vertical axis at $-1$. The curve has an absolute minimum,
of negative sign, at $z=0.85$, and then reaches a vertical asymptote 
at $z=0.97$. The second branch cuts the $z$-axis at $z=0,0.72$.
The two branches intersect each other at $z=0.15,0.69$ and at these
points $\varepsilon$ is a single-valued function. The first branch has
a vertical asymptote at $z=0.95$ whereas the second does the same
for $z$ in between $0.85$ and $0.9$.}
\end{figure}
\begin{figure}[!h]
\centerline{\hbox{\psfig{figure=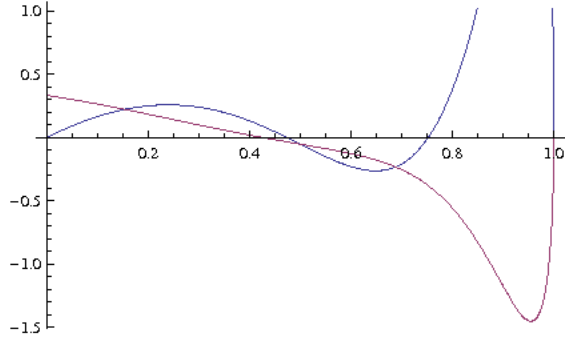,width=0.45\textwidth}}}
\caption{Two-dimensional plot of the weight function $\theta(z)$.
The first branch cuts the horizontal axis at $z=0,0.47,0.75$ and
the second cuts the horizontal axis at $z=0.42,1$. The first one
never cuts the vertical axis, whereas the second one does it
at the functional value $0.32$. The second branch has a more
pronounced absolute minimum, of negative sign, at $z=0.95$. The two
branches intersect each other at $z=0.15,0.5,0.69$, where $\theta$
is single-valued. The first branch has a vertical asymptote for
$z$ in between $0.85$ and $0.9$, whereas the second one does the
same at $z=1$.}
\end{figure}
\begin{figure}[!h]
\centerline{\hbox{\psfig{figure=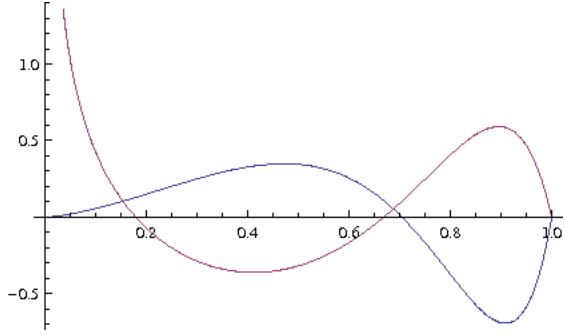,width=0.45\textwidth}}}
\caption{Two-dimensional plot of the weight function $\tau(z)=\omega(z)$.
The first branch cuts the horizontal $z$-axis at the points
$z=0,0.7,1$, and it never touches the vertical axis, whereas the
second branch cuts the $z$-axis at $z=0.18,0.67,1$, and it reaches
a vertical asymptote at $z=0.05$. The two branches intersect each other
at $z=0.15,0.69,1$, where $\tau(z)$ becomes single-valued.}
\end{figure}
\begin{figure}[!h]
\centerline{\hbox{\psfig{figure=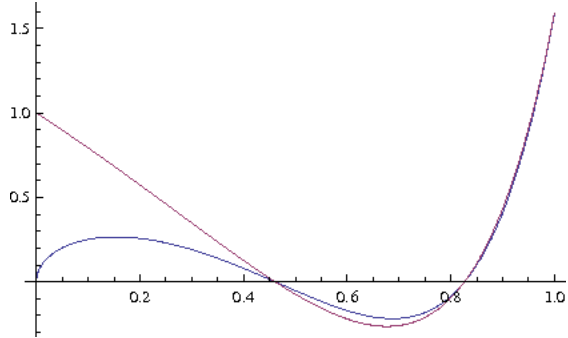,width=0.45\textwidth}}}
\caption{Two-dimensional plot of the weight functions $\pi(z)$
and ${\tilde \pi}(z)$. The $\pi(z)$ curve passes through the origin
and cuts the horizontal axis at $z=0.45,0.83$. The ${\tilde \pi}$ 
curve never passes through the origin, it cuts the vertical axis at
the functional value $1$ and it cuts the $z$-axis at $z=0.45,0.83$,
where it also intersects the $\pi(z)$ curve. Beyond the point $z=0.8$
the $\pi$ and ${\tilde \pi}$ curves become virtually indistinguishable.
At $z=1$ they both have a vertical asymptote.} 
\end{figure}
\begin{figure}[!h]
\centerline{\hbox{\psfig{figure=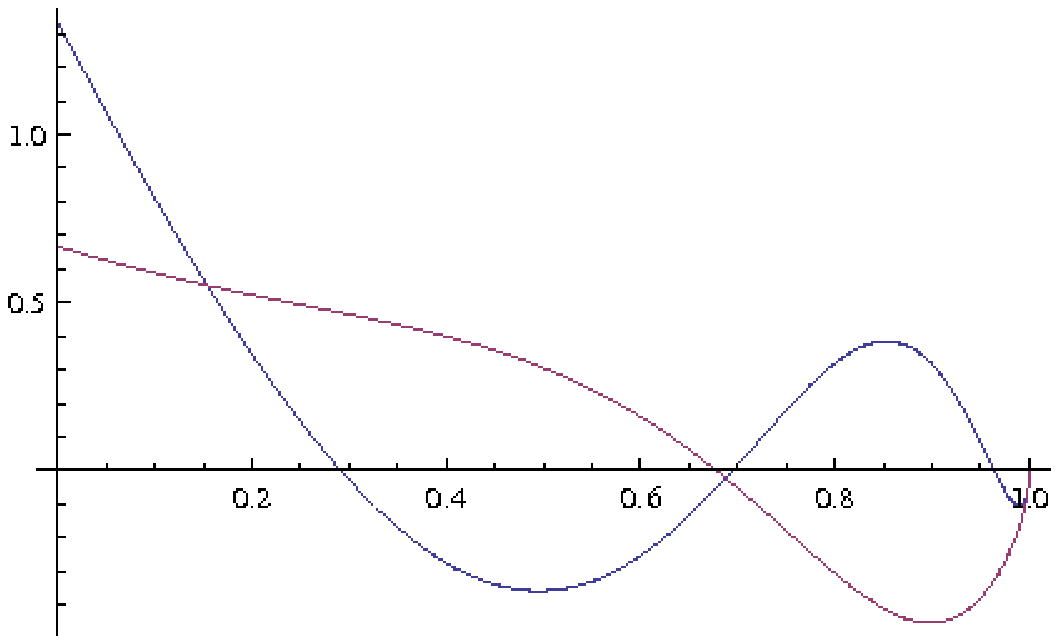,width=0.45\textwidth}}}
\caption{Two-dimensional plot of the weight function $\kappa(z)$.
The first branch cuts the $z$-axis at $z=0.29,0.7,0.97$, while the
second one cuts the $z$-axis at $z=0.68,1$. The first branch intersects
the vertical axis at the functional value $1.35$, whereas the second
one does the same at the functional value $0.67$. The two branches
intersect each other at $z=0.15,0.69$, where $\kappa$ becomes
single-valued.}
\end{figure}

As one can see, for values of $z < 0.1$, the main contribution to the
gravitino propagator results from the weight functions 
$\alpha(z),\beta(z), \tau(z)=\omega(z)$, whereas the other weight
functions are sub-dominating. By contrast, when $z \in ]0.8,1[$, the
dominating contribution to the gravitino propagator results from the
weight functions $\gamma(z),\delta(z),\varepsilon(z),\theta(z),\pi(z)$,
while the others remain sub-dominating.

\section{Concluding remarks}

Our paper has obtained the complete analytic structure of massive
gravitino propagators in de Sitter space. In Sec. VI we have plotted 
all weight functions $\alpha,\beta,\gamma,\delta,\epsilon,\theta,
\tau=\omega,\pi,\kappa$ occurring 
in the gravitino propagator (jointly with ${\tilde \pi}$) as a 
function of $z$ in a two-dimensional plot where 
$z=\cos^{2} (\mu /2R)$, $\mu$ being the geodesic distance between the
points $x$ and $x'$, and $R$ is the de Sitter radius. Although the
series (4.8) has been truncated, it remains true that Sec. VI is
the first attempt to display a supersymmetric propagator in de Sitter
via Heun functions. As we already said in Sec. I, further interest,
from the point of view of mathematical methods, arises from the
possibility to expand Heun functions in terms of hypergeometric
functions \cite{2009}. As we said before, direct implications 
of our findings on the current understanding of the 
propagation of gravitinos in de Sitter space are as follows:
there exist two ranges of values of $z$ in which the weight functions
can be divided into dominating and sub-dominating family.
In other words, when $z$ is smaller than $0.1$, the weight functions
$\alpha,\beta,\tau=\omega$ are dominating while the others are
sub-dominating. By contrast, when $z$ is very close to $1$,
the weight functions $\gamma,\delta,\varepsilon,\theta,\pi$ 
take much larger values. 
  
The plot range is between $0$ and $1$ for $z$, 
which is indeed the only admissible region, 
since the squared $\cos$ function lies always between $0$ and $1$. 
Note that the plot of $\tilde {\pi}$ is basically
nothing but the plot of the Heun function with properly defined
coefficients, and the plot of $\pi$ is $\sqrt{z}$ times the
Heun function.

The numerical analysis of Sec. VI, as we already said therein,  
has been performed by taking only
the first $10$ terms of the infinite series representing the Heun
function, by applying the Frobenius' method. If one goes on by taking 
more terms, one can get even more accurate results, but 
roughly the qualitative features remain the same. The task of plotting
Heun functions is technical but not easy, since the modern computer
packages still run into difficulties. Thus, our efforts can be
viewed as preparing the ground for a more systematic use of Heun
functions in fundamental theoretical physics. The flat-space limit
is instead a considerable simplification, since the functions $A$ 
and $C$ in (5.1) are then found to reduce to 
$A={1\over \mu}, C=-{1\over \mu}$, and the formulae in the appendix
are therefore considerably simplified.

It also remains to be seen whether the familiarity acquired with Heun
functions will prove useful in studying gravitino propagators in other
backgrounds relevant for modern high energy physics.

\appendix
\section{Explicit form of the weight functions}

The weight functions obtained in Sec. V read, explicitly,
\begin{eqnarray}
\label{alphafinalform}
&&\alpha(z)=-2mC(m^{2}+R^{-2})f^{-1}(z)\times \nonumber\\
&&\left \{ \left[(A-C)((A+C)^{2}-m^{2})
-{1\over 2}(A+C)(m^{2}+R^{-2})\right] \pi(z) 
\pm(m^{2}+R^{-2})\frac{\sqrt{z(1-z)}}{R}\pi'(z) \right \} \nonumber\\
&& -((m^{2}+R^{-2})[(A+C)^{2}-m^{2}]-4)\pi(z),
\end{eqnarray}
\begin{eqnarray}
\label{betafinalform}
&&\beta(z)=4mC(m^{2}+R^{-2})^{-1}f^{-1}(z)\times \nonumber\\
&&\left \{ \left[(A-C)((A+C)^{2}-m^{2})
-{1\over 2}(A+C)(m^{2}+R^{-2})\right] \pi(z) 
\pm(m^{2}+R^{-2})\frac{\sqrt{z(1-z)}}{R}\pi'(z) \right \} \nonumber\\
&&+2(m^{2}+R^{-2})^{-1}\left[(A+C)^{2}-m^{2}\right]\pi(z),
\end{eqnarray}
\begin{eqnarray}
\label{gammafinalform}
&&\gamma(z)=-(m^{2}+R^{-2})^{-1}\left[(A-C)^{2}
-m^{2}\right]f^{-1}(z)\times \nonumber\\
&&\left \{ \left[(A-C)((A+C)^{2}-m^{2})
-{1\over 2}(A+C)(m^{2}+R^{-2})\right] \pi(z) 
\pm(m^{2}+R^{-2})\frac{\sqrt{z(1-z)}}{R}\pi'(z) \right \} \nonumber\\
&&-2mC(m^{2}+R^{-2})^{-1}\pi(z),
\end{eqnarray}
\begin{eqnarray}
\label{gammafinalform}
&&\delta(z)=-6(m^{2}+R^{-2})^{-1}\left[(A-C)^{2}
-m^{2}\right]f^{-1}(z)\times \nonumber\\
&&\left \{ \left[(A-C)((A+C)^{2}-m^{2})
-{1\over 2}(A+C)(m^{2}+R^{-2})\right] \pi(z) 
\pm(m^{2}+R^{-2})\frac{\sqrt{z(1-z)}}{R}\pi'(z) \right \} \nonumber\\
&&+12mC(m^{2}+R^{-2})^{-1}\pi(z),
\end{eqnarray}
\begin{eqnarray}
\label{epsilonfinalform}
&&\varepsilon(z)=-(m^{2}+R^{-2})^{-1}\left[(A-C)^{2}
+\frac{2}{R^{2}}+m^{2}\right]f^{-1}(z)\times \nonumber\\
&&\left \{ \left[(A-C)((A+C)^{2}-m^{2})
-{1\over 2}(A+C)(m^{2}+R^{-2})\right] \pi(z) 
\pm(m^{2}+R^{-2})\frac{\sqrt{z(1-z)}}{R}\pi'(z) \right \} \nonumber\\
&&+2mC(m^{2}+R^{-2})^{-1}\pi(z),
\end{eqnarray}
\begin{eqnarray}
\label{thetafinalform}
&&\theta(z)=(m^{2}+R^{-2})^{-1}\left[(A-C)^{2}
-m^{2}\right]f^{-1}(z)\times \nonumber\\
&&\left \{ \left[(A-C)((A+C)^{2}-m^{2})
-{1\over 2}(A+C)(m^{2}+R^{-2})\right] \pi(z) 
\pm(m^{2}+R^{-2})\frac{\sqrt{z(1-z)}}{R}\pi'(z) \right \} \nonumber\\
&&-2mC(m^{2}+R^{-2})^{-1}\pi(z),
\end{eqnarray}
\begin{eqnarray}
\label{taumegafinalform}
&&\tau(z)=2mC(m^{2}+R^{-2})^{-1}f^{-1}(z)\times \nonumber\\
&&\left \{ \left[(A-C)((A+C)^{2}-m^{2})
-{1\over 2}(A+C)(m^{2}+R^{-2})\right] \pi(z) 
\pm(m^{2}+R^{-2})\frac{\sqrt{z(1-z)}}{R}\pi'(z) \right \} \nonumber\\
&&+(m^{2}+R^{-2})^{-1}\left[(A+C)^{2}-m^{2}\right]\pi(z) = \omega(z),
\end{eqnarray}
\begin{eqnarray}
\label{kappafinalform}
&&\kappa(z)=f^{-1}(z)\times\nonumber\\
&&\biggr \{ \left[(A-C)((A+C)^{2}-m^{2})
-{1\over 2}(A+C)(m^{2}+R^{-2})\right] \pi(z) \nonumber \\ 
& \pm & (m^{2}+R^{-2})\frac{\sqrt{z(1-z)}}{R}\pi'(z) \biggr \}.
\end{eqnarray}
These exhaust all the weight functions multiplying the 
invariant structure present in the gravitino propagator, written 
explicitly in terms of a Heun function and its derivative.

\acknowledgments

The authors are grateful to the Dipartimento di Scienze Fisiche
of Federico II University, Naples and INFN for hospitality and 
financial support. We also want to thank Ebrahim Karimi for his 
much valuable input regarding our Mathematica computations.
One of us (G.E.) dedicates this work to Maria Gabriella.


\begin{references}
\bibitem{DeWi65}
De Witt, B.S.: Dynamical Theory of Groups and Fields. 
Gordon \& Breach, New York (1965)
\bibitem{Witten98}
Witten, E.: Adv. Theor. Math. Phys. {\bf 2}, 253 (1998) 
\bibitem{Witten01}
Witten, E.: hep-th/0106109
\bibitem{raju2}
Esposito, G., Roychowdhury, R.: arXiv:0902.2098 [hep-th]
\bibitem{penrose}
Penrose, R., Rindler, W.: Spinors and Space-Time. I. 
Cambridge University Press, Cambridge (1984).
\bibitem{anguelova}
Anguelova, L., Langfelder, P.: J. High Energy Phys. JHEP {\bf 03}, 057 (2003) 
\bibitem{handbook} 
Handbook of exact solutions for ordinary 
differential equations. CRC Press, Boca Raton (1995).
\bibitem{heundiff} 
Ronveaux, A. (eds.): Heun's Differential Equations. 
Oxford University Press, Oxford (1995).
\bibitem{2009}
Sokhoyan, R.S., Melikdzanian, D.Yu., Ishkhanyan, A.M.: 
arXiv:0909.1286 [math-ph]
\bibitem{1987}
Allen, B.: Nucl. Phys. B {\bf 292}, 813 (1987)
\bibitem{allen1}
Allen, B., Jacobson, T.: Commun. Math. Phys. {\bf 103}, 669 (1986) 
\bibitem{allen2}
Allen, B., Lutken, C.A.: Commun. Math. Phys. {\bf 106}, 201 (1986) 
\bibitem{mueck} 
M\"{u}ck, W.: J. Phys. A {\bf 33}, 3021 (2000)  
\bibitem{Synge}
Synge, J.L.: Relativity: The General Theory. 
North--Holland, Amsterdam (1960)
\bibitem{abra}
Abramowitz, M., Stegun, I.A.: Handbook of Mathematical 
Functions. Dover, New York (1964)
\bibitem{erdelyi}
Erdelyi, A.: Higher Transcendental Functions. Krieger, Malabar (1981)
\bibitem{Basu}
Basu, A., Uruchurtu, L.I.: Class. Quantum Gravit. {\bf 23}, 6059 (2006) 
\bibitem{kamke} 
Kamke, E.: Differentialgleichungen, L\"{o}sungsmethoden und 
L\"{o}sungen. Vol. 1. Chelsea, New York (1974)
\bibitem{Poole36}
Poole, E.G.C.: Linear Differential Equations. Oxford University 
Press, Oxford (1936)
\bibitem{Snow52}
Snow, C. Hypergeometric and {Legendre} Functions with Applications to Integral
Equations of Potential Theory, 2nd Edition, no. 19 in Applied Mathematics
Series, National Bureau of Standards, Washington DC (1952)
\bibitem{Andrews99}
Andrews, G.E., Askey, R., Roy, R.:
Special Functions, Vol. 71 of Encyclopedia 
of Mathematics and Its Applications. Cambridge University Press, 
Cambridge (1999)
\end{references}
\end{document}